\documentstyle[prd,aps]{revtex}
\newtheorem{theorem}{Theorem}

\begin{document}
\title{The compactification of initial data on constant mean curvature time
slices in spherically symmetric spacetimes}
\author{Robert H. Gowdy}
\address{Department of Physics, Virginia Commonwealth University, PO Box\\
842000, Richmond, VA 23284-2000}
\date{July 4, 2001; gr-qc/0107016}
\maketitle

\begin{abstract}
Conformal mappings of surfaces of constant mean curvature onto compact
bounded background spaces are constructed for Minkowski space-time and for
Schwarzschild black hole spacetimes. In the black hole example, it is found
that initial data on these CMC surfaces can be regular on the compact
background space only when a certain condition is satisfied. That condition
implies that the shift vector points inward from all parts of the boundary
of the compact background. It also implies that the second fundamental form
of these surfaces can never be isotropic when black holes are present.
\end{abstract}

\section{Introduction}

With the commissioning of several gravitational wave observatories, the
calculation of the gravitational wave `signatures' of astrophysical events
has become a matter of urgency. Here, I focus on one aspect of this
calculation: The meshing of the near-field, strong curvature, part of the
calculation with the far-field region where the gravitational radiation is
actually detected.

A relatively old idea for solving this problem has recently been revived:
Use constant mean curvature (CMC) time slices, which are bounded by future
null infinity where outgoing radiation can be identified unambiguously.\cite%
{YorkPRD78,JMP80Brill,PRD96-Iriondo,PRD01CMC} The resulting constant time
surfaces are spacelike everywhere so that the standard Cauchy 3+1
formulation of the initial value data can be used. One reason that this
promising idea was abandoned for so long is the suspicion that generic
radiating spacetimes do not admit regular CMC surfaces.\cite%
{GoddardConjecture} For example, Eardley noticed that these surfaces do not
seem to be compatible with the expansions in powers of $1/r$ of fields near
future null infinity.\cite{SRCGrav-Eardley} Goddard's conjecture appears to
have been answered recently by Andersson and Iriondo who show that the
surfaces exist under quite general assumptions.\cite{pre-Andersson} I
interpret Eardley's observation as a warning against expecting CMC
time-slicing to match smoothly to the Characteristic formulation at future
null infinity. Instead, the boundary conditions that are appropriate in the
CMC description of a space-time need to be examined directly on their own,
as has been done, for example, by Friedrich.\cite%
{hyperbsFriedrich,Pune-Friedrich} The CMC description, and its
generalization, the hyperboloidal initial value problem, have begun to be
incorporated into numerical methods for solving Einstein's equations.\cite%
{PRD-Frauend1,PRD-Frauend2,CQG-huebner0,CQG-huebner1,CQG-huebner2,CQG-huebner3}

Part of the difficulty with using either the CMC description or the
hyperboloidal generalization of it is a lack of familiarity with this way of
presenting space-time. This paper provides a small step toward solving that
difficulty by analyzing two specific examples, the CMC slicings of Minkowski
space and the Schwarzschild black-hole space-time. As a preceding paper\cite%
{gowdy-ART} pointed out, one of the advantages of a CMC time-slicing is that
it promises to express all outgoing wave solutions in terms of smooth
functions on a finite conformal background. This paper finds that both the
CMC time-slicing of flat Minkowski space-time and some of the recently found
CMC time-slicings of a Schwarzschild black hole space-time\cite%
{PRD96-Iriondo} admit conformal compactifications that are simple and
regular. The necessary and sufficient condition for a CMC time-slicing of a
Schwarzschild space-time to admit a compactification in which waves are
regular functions is found to be exactly the same as the ``horizon hugging''
condition noted by A.P. Gentle {\it et. al.}\cite{PRD01CMC} The condition
has two distinct geometrical consequences: (1) The shift vector field on the
compact conformal background space is required to point inward from all
parts of the boundary, including those places where black hole event
horizons have been excised. (2) The second fundamental form is {\em not}
isotropic.

Section II of this paper reviews the hyperbolic slicing of Minkowski
space-time by surfaces of constant mean curvature and obtains the wave phase
coordinates and the compact conformal coordinates on these surfaces. Section
III uses the recently discovered CMC foliation of a Schwarzschild black hole
to obtain the wave phase coordinates and a conformal compactification for
this more complex and physically interesting case and discusses the
condition for the phase coordinate functions to be bounded so that the
compactification will be useful. Section IV discusses some of the ways in
which these specific examples may provide insights into the general case.

\section{Minkowski Space}

\subsection{Hyperbolic Slicing and the Wave Phase Coordinates}

Consider Minkowski space-time with $c=1$ units with $T$ for the standard
Minkowski time coordinate and $r$ for the spatial radius coordinate. The
asymptotically retarded time surfaces are taken to be the future-facing
spacelike hyperbolas of radius $A$ defined by the time function%
\begin{equation}
t=T-\left[ r^{2}+A^{2}\right] ^{1/2}.  \label{time}
\end{equation}%
Thus, space-time is foliated by a set of time-translated spacelike
hyperbolas, all with the same negative mean curvature, $K=-3/A$.

The previous paper\cite{gowdy-ART} began with a particular mapping of these
hyperbolas onto the three-sphere and obtained a compactification of the wave
equation. That particular mapping was not the best choice and it is better
to start with the final result and work backwards to the mapping that best
realizes it. The essential result of the previous paper was to introduce a
dimensionless space coordinate $\sigma $ such that Minkowski retarded time
takes the form%
\begin{equation}
T-r=t-\left( 3/K\right) \sigma .  \label{t-r}
\end{equation}%
This form guarantees that outgoing waves have constant phase velocity in
terms of the new variables $t,\sigma $. I will refer to $\sigma $ as the 
{\em outgoing wave phase coordinate }and denote it by $\sigma _{+}$. Notice
that outgoing waves propagate in the negative $\sigma $ direction and $%
\sigma =0$ to corresponds to infinite $R$. For the variable $\sigma $ to
have a finite range, it is necessary for the Minkowski advanced time to take
a singular form near $\sigma =0$. The form implicit in the previous paper was%
\begin{equation}
T+r=t-\left( 3/K\right) \sigma ^{-1}  \label{t+r}
\end{equation}%
from which the relation between $r$ and $\sigma $ is 
\begin{equation}
r=\frac{3}{2K}\left( \sigma -\sigma ^{-1}\right) .  \label{r(sigma)}
\end{equation}%
I will refer to $\sigma _{-}=\sigma ^{-1}$ as the {\em ingoing phase
coordinate}.

Suppose that a high frequency oscillating source of scalar waves is located
on a spherical shell at $r=r_{0}$. The wave fronts moving outward from the
shell are evenly spaced in the outgoing phase coordinate $\sigma _{+}$ while
the waves moving inward from the shell are evenly spaced in the ingoing
phase coordinate $\sigma _{-}$. If $\sigma _{+}$ takes on a finite range of
values as $r$ increases from $r_{0}$ to infinity and $\sigma _{-}$ takes on
a finite range of values as $r$ decreases from $r_{0}$, then there will, on
any given constant-$t$ surface, be a finite number of wavefronts. If a
computer program is used to construct a numerical simulation of these waves,
a finite number of mesh points will take the waves all the way to infinite $%
r $. A conformal mapping of each constant-$t$ surface onto a compact,
bounded space is one efficient way to implement such a finite mesh. Adaptive
mesh refinement is another. In any event, the key requirement for a finite
mesh to work is that the phase coordinates be {\em bounded} in their
respective directions. In the hyperbolically sliced Minkowski example, $%
\sigma _{+}$ ranges from a maximum value of $1$ at $r=0$ to a minimum value
of $0$ at infinite $r$ while $\sigma _{-}$ decreases from a value larger
than $1$ at $r=r_{0}$ to a value of $1$ at $r=0$. Thus, the phase
coordinates are indeed bounded in this simple example.

Notice that interchanging the outgoing and ingoing phase coordinates $\sigma 
$ and $\sigma ^{-1}$\ corresponds to reversing the sign of $r$ as one would
expect. The solution of Eq.~$\left( \ref{r(sigma)}\right) $ for $\sigma $
will be needed later. Of the two solutions for $\sigma $, the one that
yields positive values is%
\begin{equation}
\sigma _{+}=\sigma =\frac{1}{3}rK+\frac{1}{3}\sqrt{\left(
r^{2}K^{2}+9\right) }.  \label{minwavefn}
\end{equation}%
Similarly, the outgoing phase coordinate is%
\begin{equation}
\sigma _{-}=\sigma ^{-1}=-\frac{1}{3}rK+\frac{1}{3}\sqrt{\left(
r^{2}K^{2}+9\right) }.
\end{equation}%
It is helpful here, and later in this paper to define the scaled inverse
radius coordinate%
\begin{equation}
s=\frac{3}{\left| K\right| r}.  \label{inverserad}
\end{equation}%
The relation between the wave phase coordinates $\sigma _{\pm }$ and the
scaled inverse radius coordinate can then be put into the form%
\begin{equation}
\sigma _{\pm }=\frac{s}{\pm 1+\sqrt{1+s^{2}}}  \label{sigma-s}
\end{equation}

\subsection{Conformal Compactification}

Because waves propagating on these time-slices have just a finite number of
wavefronts, it is possible to map each slice into a finite background space
and represent the waves by regular functions on that space. One of several
reasons for insisting on a conformal mapping is that the resulting
representation is easily combined with the York conformal decomposition of
the full gravitational initial value problem.\cite{YorkPRD74,SRC78York}
Since all of the space geometries considered here are spherically symmetric,
it is a straightforward matter to construct conformal mappings between them.

The space-time metric is 
\[
ds^{2}=-dT^{2}+dr^{2}+r^{2}d\Omega ^{2} 
\]%
where $d\Omega ^{2}$ is the two-sphere metric. Use Eq.~$\left( \ref{time}%
\right) $ to express the metric in terms of the hyperbolic time parameter $t$
\begin{equation}
ds^{2}=-dt^{2}-\frac{2kr}{\sqrt{\left( r^{2}k^{2}+9\right) }}dtdr+\frac{9}{%
r^{2}k^{2}+9}dr^{2}+r^{2}d\Omega ^{2}  \label{spacetime}
\end{equation}%
so that the intrinsic metric or first fundamental form $^{3}ds^{2}$ of each
constant-$t$ surface is given by%
\begin{equation}
{}^{3}ds^{2}/A^{2}=\frac{9}{r^{2}k^{2}+9}dr^{2}+r^{2}d\Omega ^{2}.
\end{equation}%
Consider conformal mappings of each constant-$t$ surface into the unit ball $%
{B}^{3}$ in Euclidean three-space. In order to focus attention on the
boundary at infinity, use an inward directed radial coordinate $\lambda $
that is zero on the boundary and takes the value $\lambda =1$ at the center.
The metric on ${B}^{3}$ then takes the form 
\[
d\ell ^{2}=d\lambda ^{2}+\left( 1-\lambda \right) ^{2}d\Omega ^{2} 
\]%
\ For the mapping defined by $r=f\left( \lambda \right) $ to be conformal,
these metrics must be related by%
\[
{}^{3}ds^{2}=\Phi ^{4}d\ell ^{2}. 
\]%
This condition reduces to a first order ordinary differential equation for $%
f,$ which is readily integrated. The solutions that satisfy $f\left(
1\right) =0,f\left( 0\right) =\infty $ are most easily expressed in terms of
the inverse radius coordinate $s$ defined by Eq.~$\left( \ref{inverserad}%
\right) $. \ The relationship between $s$ and the conformal coordinate $%
\lambda $ is 
\begin{equation}
s=\frac{\lambda \left( 2-\lambda \right) }{2\left( 1-\lambda \right) }
\label{s-lambda}
\end{equation}%
and the conformal factor becomes

\begin{equation}
\Phi ^{2}=-\left( 6/K\right) \left[ \lambda \left( 2-\lambda \right) \right]
^{-1}
\end{equation}%
The resulting form of the space-time metric can be obtained by expressing
the space-time metric (See Eq.~$\left( \ref{spacetime}\right) $.) in terms
of $\lambda =x^{1}$ with the result%
\begin{equation}
ds^{2}=-dt^{2}-\frac{24}{K}\frac{1-\lambda }{\lambda ^{2}\left( 2-\lambda
\right) ^{2}}dtd\lambda +\frac{36}{K^{2}}\frac{1}{\lambda ^{2}\left(
2-\lambda \right) ^{2}}\left( d\lambda ^{2}+\left( 1-\lambda \right)
^{2}d\Omega ^{2}\right)  \label{flatconform}
\end{equation}

The main reason for suspecting that the conformal map would be useful is the
bounded nature of the wave phase coordinates. Thus it is not too surprising
that these are simply related to the conformal coordinate. From equations $%
\left( \ref{sigma-s}\right) $ and $\left( \ref{s-lambda}\right) $ the
relation is%
\[
\sigma _{+}=\frac{\lambda }{2-\lambda },\qquad \sigma _{-}=\frac{2-\lambda }{%
\lambda } 
\]

\subsection{3+1 Initial Data for Flat Space-time}

The usual procedure for generating a curved space-time begins with Cauchy
initial data on a spacelike surface. Thus, it is useful to summarize the 3+1
initial data that corresponds to this hyperbolic slicing time-slicing of
flat space-time. The Arnowitt, Deser, and Misner 3+1 notation will be used.%
\cite{ADM62Witten} This data consists of the second fundamental form or
extrinsic curvature $K_{ij}$ split into its trace $K$, and its trace-free
part $A_{ij}=K_{ij}-\left( 1/3\right) Kg_{ij}$, the three dimensional metric
tensor $g_{ij}$, the lapse function $N$ and the shift vector $\vec{N}$.
Solutions to the initial value constraints are usually found by using the
York conformal rescaling procedure which expresses the actual space metric
in terms of a conformal background metric $\tilde{g}_{ij}$ and solves the
Hamiltonian constraint for the conformal factor\cite{YorkPRD74,SRC78York}.

The second fundamental form or extrinsic curvature of a hyperbolic time
slice is readily found to be%
\[
K_{ij}=-A^{-1}g_{ij}=\frac{K}{3}g_{ij} 
\]%
where $g_{ij}$ is the three-metric on the slice. The trace-free part $A_{ij}$
of $K_{ij}$ is zero in this case. The background space metric $\tilde{g}%
_{ij} $ is simply the metric on Euclidean space.

The lapse and shift functions, $N$ and $\vec{N}$, can be obtained from Eq.~$%
\left( \ref{flatconform}\right) $, which yields the radial shift form
component 
\begin{equation}
N_{1}=-\frac{24}{K}\frac{1-\lambda }{\lambda ^{2}\left( 2-\lambda \right)
^{2}}  \label{coshift}
\end{equation}%
and radial shift vector component%
\begin{equation}
N^{1}=g^{11}N_{1}=\Phi ^{-4}N_{1}=-\frac{2}{3}K\left( 1-\lambda \right)
\label{mshift}
\end{equation}%
The lapse function is then%
\[
N=\sqrt{1+N_{1}N^{1}}=\sqrt{1+16\frac{\left( 1-\lambda \right) ^{2}}{\lambda
^{2}\left( 2-\lambda \right) ^{2}}.} 
\]%
Notice that it is the well-behaved shift vector given by Eq.~$\left( \ref%
{mshift}\right) $ that actually enters the dynamical equations through the
Lie derivative of the background metric $\tilde{g}$ and not the singular
shift form $N_{i}$ described by Eq.~$\left( \ref{coshift}\right) $. The
positive value of the radial shift vector component corresponds to a {\em %
shift vector that points inward} as one would expect because the local
inertial frames of these hyperbolic surfaces are expanding outward.

The lapse function, as a scalar quantity on the three-surface, diverges at
the boundary of the three-ball just as it diverges at infinite $r$. A
conformally rescaled lapse function that remains finite everywhere on the
three-ball can be defined as follows:%
\[
\tilde{N}=\Phi ^{-2}N=-\frac{K}{6}\sqrt{\left( 2-\lambda \right) ^{2}\lambda
^{2}+16\left( 1-\lambda \right) ^{2}}. 
\]%
This rescaled lapse corresponds to applying the conformal factor, $\Phi
^{4}, $ to the entire space-time metric. This procedure is similar to the
Penrose compactification procedure but with one important difference: Here
the conformal factor is time independent and the resulting boundary at
future null infinity is a symmetric $S^{2}\times {R}$ cylinder.

\section{The Schwarzschild Black Hole}

\subsection{CMC slicings and the naked shift-reversal condition}

A remarkably simple static form of the Schwarzschild metric, found by
Iriondo, Malec, and O'Murchadha\cite{PRD96-Iriondo} and analyzed recently by
Adrian P. Gentle {\it et. al.}\cite{PRD01CMC}{\it \ }displays the most
general spherically symmetric CMC time-slicing of a black hole space-time.
The metric tensor takes the form 
\begin{equation}
ds^{2}=-\left( 1-2m/r\right)
dt^{2}+2vP^{-1/2}dtdr+r^{4}P^{-1}dr^{2}+r^{2}d\Omega ^{2}  \label{CMCschw}
\end{equation}%
where $v,P$ are polynomials%
\begin{equation}
P=v^{2}+\left( 1-2m/r\right) r^{4}  \label{polynomial}
\end{equation}%
\begin{equation}
v=Kr^{3}/3-H  \label{v}
\end{equation}%
$K$ is the mean curvature, $m$ is the mass of the black hole, and $H$ is an
integration constant. The radius coordinate $r$ that is used here is just
the usual two-sphere area radius. A straightforward computation of the
second fundamental form of the $t=$ constant surfaces yields the components 
\[
K_{11}=g_{11}\left( K/3+2H/r^{3}\right) ,\qquad K_{22}=g_{22}\left(
K/3-H/r^{3}\right) ,\qquad K_{33}=g_{33}\left( K/3-H/r^{3}\right) 
\]%
where the coordinates are labeled in the usual way with $r=x^{1},$ $\theta
=x^{2},\varphi =x^{3}$. The integration constant $H$ measures the
anisotropic, trace-free curvature. Since these surfaces are conformally
flat, the case $H=0$ corresponds to initial data that is completely
isotropic everywhere and only this case leads to surfaces that resemble
hyperbolas at infinity. The value of $H$ is determined by boundary
conditions at the `center' of the system. For the $m=0$ case, regularity at $%
r=0$ requires that $H$ vanish. For $m>0$, the value of $H$ must be adjusted
to produce the best foliation of the black hole near its horizon.\cite%
{PRD01CMC}

The shift vector field, $\vec{N}$ describes the three-velocity of
constant-coordinate lines relative to the local inertial frame defined by
the time-slice. From the form of the space time metric given by Eq.~$\left( %
\ref{CMCschw}\right) $, the radial shift vector component is%
\begin{equation}
N^{1}=r^{-4}v\sqrt{P}=r^{-4}\left( Kr^{3}/3-H\right) \sqrt{P}.  \label{shift}
\end{equation}%
So long as $K$ is negative and $r$ is sufficiently large, the shift vector
field points inward, just as in the case of hyperbolically sliced Minkowski
space. An inward shift vector means that the local inertial frames of the
surface are expanding outward. However, unlike the Minkowski space case, the
local inertial frames that are sufficiently near a black hole can be falling
inward instead of expanding outward. The shift vector then reverses. For the
reversal to happen, $H$ and $K$ must have the same sign. From the above
expression, this reversal happens at%
\[
r_{0}=\left( \frac{3H}{K}\right) ^{1/3}. 
\]%
For the reversal to happen outside the event horizon, we need the condition%
\[
r_{0}>2m. 
\]%
We will see this same combination of requirements again later, so I will
give it a name: the {\em naked shift reversal} {\em condition}. The
condition implies%
\[
\left| H\right| >\frac{8}{3}m^{3}\left| K\right| , 
\]%
which for negative $K$ , becomes%
\[
H<H_{+} 
\]%
where%
\[
H_{+}=\frac{8}{3}m^{3}K 
\]%
and is identical to one used by A.P. Gentle {\it et.al}. to describe the CMC
foliations that they characterize as useful. Their paper points out the
presence of the shift reversal outside the event horizon in these CMC
foliations.\cite{PRD01CMC}

\subsection{Bounded wave phase coordinates}

As in the case of hyperbolically sliced Minkowski space, outgoing waves can
be represented on CMC time-slices as smooth functions of an outgoing phase
coordinate $\sigma _{+}\left( r\right) $ and an ingoing phase coordinate $%
\sigma _{-}\left( r\right) $\ . If these phase coordinates are {\em bounded}
in their respective directions, then a conformal mapping can represent waves
as regular functions on a compact background space. Thus, our first task is
to seek the phase coordinates. As in the Minkowski space example, take $%
U_{\pm }=t-\frac{3}{K}\sigma _{\pm }\left( r\right) $ to be null coordinates
on space-time so that ingoing and outgoing waves move at constant velocities 
$d\sigma _{\pm }/dt$ in their respective directions. The corresponding
general requirement is the null condition $dU\cdot dU=0$ which, for the CMC
sliced Schwarzschild metric, yields%
\begin{equation}
\left( 1-\frac{2m}{r}\right) \frac{3}{K}\frac{d\sigma _{\pm }}{dr}%
=vP^{-1/2}+b_{\pm }  \label{CMCnull}
\end{equation}%
where $b=\pm 1$. The choice of $b$ determines whether $\sigma _{\pm }$ is
the outgoing phase coordinate or the ingoing phase coordinate.

The outgoing phase coordinate $\sigma _{+}$ must be bounded as $r$ increases
which means that $d\sigma _{+}/dr$ should go to zero in that limit. The
choices that meet this requirement are:%
\begin{equation}
b_{+}=\left\{ 
\begin{array}{ccc}
-1 & \text{for} & K>0 \\ 
+1 & \text{for} & K<0%
\end{array}%
\right.  \label{bplus}
\end{equation}%
The opposite choices must then correspond to the ingoing phase coordinate so
that%
\begin{equation}
b_{-}=\left\{ 
\begin{array}{ccc}
+1 & \text{for} & K>0 \\ 
-1 & \text{for} & K<0%
\end{array}%
\right.  \label{bminus}
\end{equation}

The ingoing phase coordinate $\sigma _{-}$ and therefore its derivative $%
d\sigma _{-}/dr$ needs to be bounded as $r$ decreases past the Schwarzschild
limit at $r=2m$. Write Eq.~$\left( \ref{CMCnull}\right) $ in the form%
\[
\left( 1-2m/r\right) \frac{3}{K}\frac{d\sigma _{-}}{dr}=P^{-1/2}\left(
v+b_{-}\sqrt{P}\right) 
\]%
and note that%
\[
\left( -b_{-}v+\sqrt{P}\right) \left( b_{-}v+\sqrt{P}\right) =P-v^{2}=\left(
1-2m/r\right) r^{4}
\]%
so that the expression for the derivative becomes%
\[
\frac{d\sigma _{-}}{dr}=\left( K/3\right) \frac{b_{-}}{\sqrt{P}}\frac{r^{4}}{%
-b_{-}v+\sqrt{P}}
\]%
Because the polynomial $P$ is positive for all $r\geq 2m$, this expression
remains finite at $r=2m$ if and only if $b_{-}v<0$. From Eq.~$\left( \ref%
{bminus}\right) $, the necessary and sufficient condition for $b_{-}v<0$ is
that $v$ \ and $K$ have opposite signs at $r=2m$. From the above expression,
it can be seen that $\frac{d\sigma _{-}}{dr}$ then has bounded magnitude and
is bounded away from zero for any values of $r$ in the range $2m\leq r<R$
where $R$ is any number larger than $2m$. Thus, a bounded ingoing phase
coordinate $\sigma _{-}$ can extend across the event horizon at $r-2m$ if
and only if $v$ and $K$ have opposite signs at $r=2m$,

Now recall that $v$ is related to the radial shift vector component $N^{1}$
by Eq.~$\left( \ref{shift}\right) $ and, from Eq.$\left( \ref{v}\right) $
has the {\em same sign} as $K$ for sufficiently large values of $r$. The
condition that $v$ and $K$ have opposite signs at $r=2m$ is then equivalent
to requiring a naked shift reversal and the following result has been
established:

\begin{theorem}
The naked shift reversal condition is necessary and sufficient for the
existence of bounded wave phase coordinates on a constant mean curvature
slice in the exterior region of a Schwarzschild space-time. The resulting
phase coordinates are monotonic functions of the luminosity radius $r$.
\end{theorem}

As an example of the value of considering a specific example, notice that
the naked shift reversal condition {\em rules out} the simplifying
assumption, $K_{ij}=\left( 1/3\right) Kg_{ij}$ that underlies many of the
results obtained by Friedrich.\cite{hyperbsFriedrich,Pune-Friedrich} For
these black hole space times, that assumption $\left( H=0\right) $ leads to
constant-time surfaces on which the ingoing phase coordinate is not
bounded.\ Thus the $K_{ij}=\left( 1/3\right) Kg_{ij}$ assumption contradicts
the assumption that these constant-time surfaces can be mapped into a
compact manifold where waves may be described by regular functions.

\subsection{Conformal Compactification}

If the wave phase coordinates exist and behave properly, we expect that it
should be possible to map all of the outgoing wave action in one of these
CMC sliced spacetimes onto ${\rm B}^{3}$. Thus, we seek a function $\lambda
\left( r\right) $ that corresponds to a conformal map of the CMC spatial
metric%
\[
^{3}ds^{2}=\frac{r^{4}}{P}dr^{2}+r^{2}d\Omega ^{2} 
\]%
into the metric%
\[
\Phi ^{4}\left( d\lambda ^{2}+\left( 1-\lambda \right) ^{2}d\Omega
^{2}\right) 
\]%
The resulting condition on $\lambda $ is%
\begin{equation}
\frac{d\lambda }{1-\lambda }=-\frac{rdr}{\sqrt{\left( H-\frac{1}{3}%
Kr^{3}\right) ^{2}+\left( 1-\frac{2m}{r}\right) r^{4}}}  \label{CMC-lambda}
\end{equation}%
and can be integrated to 
\begin{equation}
\lambda =1-e^{-F\left( m,H,K,s\right) }  \label{lambda-1}
\end{equation}%
where $s$ is the inverse radius coordinate given by Eq.~$\left( \ref%
{inverserad}\right) $ and $F$ is the integral%
\begin{equation}
F\left( m,H,K,s\right) =\int_{0}^{s}\frac{dx}{\sqrt{\left( 1+\left(
1/9\right) HK^{2}x^{3}\right) ^{2}+x^{2}\left( 1+\frac{2mKx}{3}\right) }}.
\label{lambda-2}
\end{equation}

The coordinate transformation from the luminosity radius $r$ to the
three-ball coordinate $\lambda $ is only available in an implicit form,
given by equations $\left( \ref{lambda-1}\right) $ and $\left( \ref{lambda-2}%
\right) $. However, it is easy to calculate the derivative $d\lambda /dr$
from this form and then the combination $P^{-1/2}\frac{dr}{d\lambda }$ that
is needed to transform the metric. The result is found to be amazingly
simple.%
\[
P^{-1/2}\frac{dr}{d\lambda }=-\frac{1}{r}\frac{1}{1-\lambda } 
\]%
The resulting form of the space-time metric is then%
\begin{equation}
ds^{2}=-\left( 1-2m/r\right) dt^{2}-2\frac{Kr^{2}/3-H/r}{1-\lambda }%
dtd\lambda +\frac{r^{2}}{\left( 1-\lambda \right) ^{2}}d\lambda
^{2}+r^{2}d\Omega ^{2}  \label{CMC-Schw-compact}
\end{equation}%
and the conformal factor is%
\[
\Phi ^{2}=\frac{r}{1-\lambda } 
\]%
while the shift vector (for the coordinate $\lambda $) is%
\[
N^{1}=-\left( K/3-H/r^{3}\right) \left( 1-\lambda \right) . 
\]%
Just as for the Minkowski space example, we discover a compactified solution
that is built entirely from rational polynomials.

\subsection{Relation between phase coordinates}

In the Minkowski case it was seen that the phase coordinates $\sigma _{\pm }$
and the coordinate $\lambda $ on the compact conformal background are simply
related. In the Schwarzschild geometry, we do not have the luxury of
analytic forms for $\sigma _{\pm }$ but we can seek relationships directly
from the differential equations that they satisfy. From equations $\left( %
\ref{CMCnull}\right) $ and $\left( \ref{CMC-lambda}\right) $ for the case $%
K<0,$ 
\begin{equation}
\left( 1-\frac{2m}{r}\right) \frac{3}{K}\frac{d\sigma _{+}}{dr}=vP^{-1/2}+1
\label{a}
\end{equation}%
\begin{equation}
\left( 1-\frac{2m}{r}\right) \frac{3}{K}\frac{d\sigma _{-}}{dr}=vP^{-1/2}-1.
\label{b}
\end{equation}%
Subtract Eq.~$\left( \ref{b}\right) $ from Eq.~$\left( \ref{a}\right) $ and
obtain a simple relation between the ingoing and outgoing phase coordinates 
\[
\frac{3}{K}\left( \sigma _{+}-\sigma _{-}\right) =r_{0}\ln \left(
r_{0}-2m\right) -r\ln \left( r-2m\right) 
\]%
where $r_{0}$ can be chosen to have any value greater than $2m$.

\subsection{Location of the event horizon in the three-ball}

The conformal coordinate $\lambda $ ranges from a value of zero at the outer
boundary of the three-ball to a value of $\lambda =1$ at the center. The
event horizon at $r=2m$ is located at the conformal coordinate value%
\[
\lambda _{h}\left( H,K,m\right) =1-e^{-F\left( 1,H,K,3/\left( 2m\left|
K\right| \right) \right) } 
\]%
For the values $H=-1.25,K=-0.1,m=1$ that A.P. Gentle {\it et.al}.\cite%
{PRD01CMC} use as an illustration of CMC foliation, the horizon is located at%
\[
\lambda _{h}\left( -1.25,-0.1,1\right) =0.987\,71. 
\]%
The coordinate radius $1-\lambda $ of the horizon surface would then be $%
\allowbreak 0.012\,29$ and is thus quite small in this picture. A black hole
with twice the mass and the same CMC time parameters, would have its event
horizon at%
\[
\lambda _{h}\left( -1.25,-0.1,2\right) =\allowbreak 0.948\,78. 
\]%
Thus, it would have much more than twice the coordinate size in this
description. The horizon size is not very sensitive to the anisotropic
curvature parameter $H$. For example, with $m=2$ the naked shift reversal
condition requires that $H$ be less than $-2.133$. Changing the value of $H$
from $-1.25$ to $-3$ results in only a small change in the location of the
event horizon.%
\[
\lambda _{h}\left( -3,-0.1,2\right) =\allowbreak 0.949\,4. 
\]%
The scale of the representation is set by the mean curvature parameter $K$.
However, even with the choice $K=-1$, a unit mass black hole would still be
quite close to the origin of the three-ball with%
\[
\lambda _{h}\left( -3,-1,1\right) =\allowbreak 0.871\,13. 
\]

\section{Discussion}

Here, the constant mean-curvature initial conditions for Minkowski space and
the Schwarzschild black hole space-time, have been described in terms of
regular functions on the unit three-ball. The condition that waves also be
described by regular functions of the compact coordinates has been found to
restrict these initial conditions in a natural way: If one thinks of the
region outside the black hole event horizon as the active region, then the
shift vector at the inner and outer boundaries of that region must point
into the active region. A somewhat less obvious result is that requiring
waves to be described by regular functions of the compact coordinates rules
out the assumption that the second fundamental form is isotropic. This
simplifying assumption cannot be used to analyze the initial value problem
for spacetimes which contain black holes.

Both the Minkowski space and the Schwarzschild black hole metrics display a
curious feature when expressed in compact CMC coordinates. In both cases,
the metric functions are built from rational polynomials. In the
Schwarzschild case, the polynomial fractions involve both the compact radius
coordinate $\lambda $ and the area-radius $r\left( \lambda \right) $. It is
generally understood that the metric functions are not analytic functions of 
$1/r$ so that expansions in powers of $1/r$ are, at best, only
asymptotically convergent. Here, this nonanalyticity is concentrated in the
area-radius function $r\left( \lambda \right) $. It is a straightforward
exercise to use Equations \ref{lambda-1} and \ref{lambda-2} to expand $1/r$
in powers of $\lambda $. One learns that the series deviates from the
Minkowski space solution (See Equations \ref{inverserad} and \ref{s-lambda})
for $1/r$ only in fourth order and above and is sharply divergent for all $%
r<15m$.

\end{document}